\documentclass[twocolumn,english,amsart,showpacs,preprintnumbers,amsmath,amssymb,floatfix,nofootinbib]{revtex4-1}

\usepackage{enumitem}
\usepackage[T1]{fontenc}
\usepackage{adjustbox}
\usepackage[latin9,utf8]{inputenc}
\usepackage{xcolor}
\usepackage{array}
\usepackage{amstext}
\usepackage{graphicx}
\usepackage{epstopdf}
\usepackage{esint}
\usepackage{float}
\usepackage{tabularx}
\usepackage{dcolumn}
\usepackage{bm}
\usepackage{subfigure}
\usepackage{color}
\usepackage{float}
\usepackage{multirow}
\usepackage{xspace}
\usepackage{booktabs}
\usepackage{amsmath}
\usepackage{hhline,multirow,tabularx}
\usepackage{latexsym,amssymb}
\usepackage[mathscr]{eucal}
\usepackage{rotating}
\usepackage{appendix}
\usepackage{ulem}
\usepackage{tikz,xcolor}
\usepackage{url}
\usepackage{placeins}
\usepackage[colorlinks = true,
linkcolor = blue,
urlcolor  = blue,
citecolor = blue,
anchorcolor = blue,
linktocpage]{hyperref}
\definecolor{lime}{HTML}{A6CE39}
\DeclareRobustCommand{\orcidicon}{
	\begin{tikzpicture}
		\draw[lime, fill=lime] (0,0)
		circle [radius=0.16]
		node[white] {{\fontfamily{qag}\selectfont \tiny ID}};
		\draw[white, fill=white] (-0.0625,0.095)
		circle [radius=0.007];
	\end{tikzpicture}
	\hspace{-2mm}
}
\foreach \x in {A, ..., Z}{
	\expandafter\xdef\csname orcid\x\endcsname{\noexpand\href{https://orcid.org/\csname orcidauthor\x\endcsname}{\noexpand\orcidicon}}
}

\newcommand{\x}{\ensuremath{x}\xspace}

\makeatletter

\usepackage{xcolor}
\usepackage[pagewise]{lineno}
\definecolor{darkred}{rgb}{0.6,0,0}
\begin{document}


\title{
NNLO Determination of Polarized Parton Distribution Functions
with Higher-Twist and Target-Mass Corrections
}

\author{Javad Shahrzad\orcidD{}
}  
\email{Javadshahrzad@semnan.ac.ir}
\affiliation{Faculty of Physics, Semnan University, P. O. Box 35131-19111, Semnan, Iran}

\author{Elliot Leader\orcidB{}}
\email{e.leader@imperial.ac.uk}
\affiliation{Imperial College,
Prince Consort Road, London SW7 2BW, England
}%

\author{Ali Khorramian\orcidA{}}
\email{Khorramiana@semnan.ac.ir}
\affiliation{Faculty of Physics, Semnan University, P. O. Box 35131-19111, Semnan, Iran~}%


\begin{abstract}
We present a next-to-next-to-leading order (NNLO) QCD
analysis of polarized parton distribution functions (PDFs)
based on the world data set of inclusive polarized
deep-inelastic scattering (DIS). The resulting PDF set,
denoted as \texttt{KLS26}, includes a consistent treatment
of target-mass corrections (TMCs), higher-twist (HT)
contributions, and positivity constraints within a fully
NNLO framework. To estimate the residual theoretical uncertainty
associated with nonperturbative power corrections, both
additive and multiplicative HT parametrizations are
considered. Although these two approaches lead to
different higher-twist coefficients, the current results yield very
similar fit qualities and compatible polarized PDFs within
uncertainties.
The impact of NNLO corrections is assessed through a
comparison with the KLSS21 polarized PDF
determination at NLO, showing the most visible changes in
the polarized strange-quark distribution, together with
moderate modifications in the gluon sector and smaller
effects in the up- and down-quark helicity distributions. Additional fits treating the nonsinglet axial charges as free parameters, together with different higher-twist implementations, induce small but flavor-dependent modifications in the PDFs, with the most noticeable effects in the polarized strange-quark distribution and moderate sensitivity in the gluon. The \texttt{KLS26} set establishes an updated determination of helicity-dependent parton distributions, enabling future precision studies of nucleon spin structure.

\end{abstract}

\maketitle

\section{Introduction}

The helicity-dependent (polarized) parton distribution functions (PDFs) describe the difference between parton number densities with spin aligned parallel and anti-parallel to the spin of the proton~\cite{Leader:2010rb}. These distributions encode fundamental information about the spin structure of the nucleon and play a central role in understanding the decomposition of the proton spin within Quantum Chromodynamics (QCD)~\cite{deFlorian:2008mr}.

\begin{figure}[ht]
\centering
\includegraphics[width=0.3\textwidth]{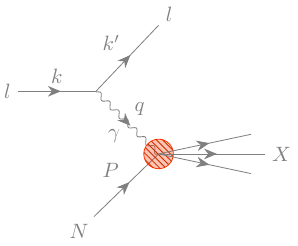}
\caption{Feynman diagram for inelastic lepton--nucleon scattering,
$lN \rightarrow lX$.}
\label{fig:DIS}
\end{figure}

The polarized parton distribution functions are measured via deep
inelastic lepton--hadron scattering,
\begin{equation}
l + N \rightarrow l + X\,,
\end{equation}
in which the incoming lepton is longitudinally polarized
($\rightarrow$), and the nucleon target, in the case we are interested in,
is polarized either along ($\Rightarrow$) or opposite ($\Leftarrow$)
to the momentum of the lepton beam (see Fig.~\ref{fig:DIS}).

We consider the reaction in the laboratory frame, in which the nucleon is at rest. The initial and final lepton four-momenta are written as

\begin{equation}
k^\mu = (E,\mathbf{k})\,,
\qquad
k'^\mu = (E',\mathbf{k}')\,,
\label{eq:k}
\end{equation}
and for the initial nucleon four-momentum
\begin{equation}
P^\mu = (M,\mathbf{0})\,,
\label{eq:P}
\end{equation}
where $M$ denotes the nucleon mass.

The difference between the differential cross sections corresponding to parallel and antiparallel configurations of the lepton and nucleon spins is expressed through two scalar spin-dependent structure functions, $g_1$ and $g_2$:
\begin{eqnarray}
    \frac{d^{2}\sigma^{\substack{\rightarrow\\\Rightarrow}}}{d\Omega\, dE'}&-&\frac{d^{2}\sigma^{\substack{\rightarrow\\\Leftarrow}}}{d\Omega\, dE'}= \nonumber\\
    &-&\frac{4\alpha^{2}E'}{Q^{2}EM\nu}
    \left[
    (E+E'\cos\theta)\,g_{1}
    -
    2xMg_{2}
    \right],
    \label{eq:polxsec}
\end{eqnarray}
where $\theta$ is the laboratory scattering angle of the lepton.

In Eq.~(\ref{eq:polxsec}), $\alpha$ is the fine-structure constant, $\nu = E-E'$, $Q^{2}=-q^{2}$,
where $q=k-k'$ and the structure functions depend on the variables $x$ and $Q^{2}$,
where
$
x \equiv x_{\rm Bjorken}
=
\frac{Q^{2}}{2P \cdot q}
=
\frac{Q^{2}}{2M\nu}
$.

In all experiments up to the present, with polarized beam and target,
the kinematic region studied is very limited compared to the unpolarized
case, and it is thus considered safe to neglect the $g_{2}$ term in
Eq.~(\ref{eq:polxsec}). Hence, present and previous experiments of this
type are considered simply as measurements of the structure function
$g_{1}(x,Q^{2})$ which, as will be seen shortly, can be expressed directly in terms of the polarized parton densities.

Over the past decades, extensive theoretical and phenomenological efforts have been devoted to extracting polarized PDFs from polarized deep inelastic scattering (DIS), semi-inclusive DIS (SIDIS), and polarized proton--proton collision data. Several global QCD analyses have been performed with different methodologies, parametrizations, and perturbative accuracies.

Representative examples include the DSSV analyses~\cite{deFlorian:2009vb,deFlorian:2014yva}, the NNPDFpol framework~\cite{Nocera:2014gqa}, the JAM collaboration analyses~\cite{Sato:2016tuz,Etheridge:2022yer}, the AAC approach~\cite{Hirai:2003pm}, the BB analyses~\cite{Blumlein:2010rn}, and the LSS framework~\cite{Leader:2014uua,LSS11} and its later extensions, including the KLSS21 analyses~\cite{Khorramian:2020gkr}.

In parallel, several analyses based on Jacobi polynomial techniques, as well as alternative fitting approaches, have been carried out by our group and others~\cite{Salimi-Amiri:2018had,TaheriMonfared:2014var,Jimenez-Delgado:2013boa,Khorramian:2010qa,AtashbarTehrani:2007odq,Khorramian:2004ih,Arbabifar:2013tma}. Despite these advances, polarized PDFs remain considerably less constrained than their unpolarized counterparts, particularly in the strange quark and gluon sectors.

More recently, next-to-next-to-leading order (NNLO)-accurate extractions such as BDSSV, MAPPDFpol1.0, and NNPDFpol2.0 have significantly improved the theoretical precision of helicity PDF determinations~\cite{Borsa:2024mss,Bertone:2024taw,Cruz-Martinez:2025ahf}. Although recent NNLO analyses incorporate a broad range of experimental data, inclusive polarized DIS remains the cleanest process for studying the spin-dependent structure of the nucleon due to the absence of fragmentation-function uncertainties. It therefore provides a controlled environment for disentangling perturbative and nonperturbative QCD effects. This is particularly relevant for the present analysis, where special emphasis is placed on the polarized strange quark distribution and on the impact of additive and multiplicative higher-twist contributions. Within this framework, inclusive DIS allows a transparent and consistent NNLO determination.

Inclusive and semi-inclusive polarized DIS measurements primarily drive current knowledge of spin-dependent partonic structures. Inclusive DIS primarily constrains the flavor combinations
\(
\Delta q(x,Q^2)+\Delta \bar{q}(x,Q^2),
\)
while SIDIS data, combined with fragmentation functions, provide additional flavor separation. Nevertheless, significant uncertainties remain, especially for the polarized strange quark distribution and the gluon helicity distribution.

A long-standing issue in polarized QCD analyses is the so-called strange quark polarization puzzle. Inclusive DIS analyses tend to favor a negative strange quark polarization,
\(
\Delta s(x,Q^2)+\Delta \bar{s}(x,Q^2),
\)
even when flexible parametrizations are employed. In contrast, global analyses including SIDIS data often indicate a positive contribution in the intermediate-\(x\) region, 
while remaining negative at smaller \(x\). This tension is strongly influenced by the choice of kaon fragmentation functions and by assumptions regarding SU(3) flavor symmetry in axial charges, as will be discussed in Section~\ref{subsec:free_axial_charges}.

The determination of the polarized strange quark distribution is known to be particularly sensitive to theoretical assumptions entering global analyses. In SIDIS-based extractions, the resulting strange quark polarization depends strongly on the choice of kaon fragmentation functions~\cite{LSS11}. Furthermore, the degree of SU(3) flavor symmetry breaking in the axial charges can significantly affect the extracted shape and normalization of
\(
\Delta s(x,Q^2)+\Delta \bar{s}(x,Q^2).
\)
Previous investigations, including our own NLO analyses~\cite{Khorramian:2020gkr}, have demonstrated the importance of these effects and motivated a more precise study within an NNLO framework.

Another important ingredient in polarized DIS analyses is the treatment of higher-twist (HT) effects, which become increasingly relevant at moderate and large values of \(x\) and at relatively low \(Q^2\). Since a significant fraction of polarized DIS data resides in this kinematic region, a consistent treatment of HT contributions is essential for a reliable extraction of helicity PDFs. In this work, we systematically investigate higher-twist effects for both proton and neutron data, considering both additive and multiplicative parametrizations. This allows us to assess the stability of the extracted polarized PDFs under different modeling assumptions and to quantify the associated theoretical uncertainties.

Positivity constraints also play an important role in polarized PDF extractions. At leading order, helicity PDFs satisfy the bound
\begin{equation}
|\Delta f_i(x,Q^2)| \leq f_i(x,Q^2)\,,
\label{eq:positivity}
\end{equation}
reflecting the positivity of physical cross sections at leading order. However, beyond leading order, PDFs become factorization-scheme dependent, and strict point-by-point positivity is no longer guaranteed. Instead, the relevant requirement is the positivity of physical observables. In practice, positivity remains a powerful phenomenological constraint, particularly in poorly constrained regions such as the polarized gluon and strange quark sectors. Previous studies have demonstrated that enforcing positivity significantly stabilizes global fits and reduces unphysical behavior in helicity PDFs~\cite{Leader:2001gr,Leader:2005ci,Altarelli:1998gn}.

The perturbative stability of positivity constraints in polarized QCD analysis has been investigated in recent study~\cite{Candido:2023}. This analysis demonstrated that polarized PDFs evolved in the \(\overline{\mathrm{MS}}\) factorization scheme remain effectively positive throughout the perturbative region, provided that the evolution originates from a physically positive scheme and that the corresponding perturbative transformations are consistently implemented. Possible departures from positivity may occur only at sufficiently low scales where perturbation theory itself becomes unreliable. For the kinematic region relevant to modern global analyses,
\(Q \gtrsim \mathcal{O}(1\text{--}2~\mathrm{GeV})\),
positivity remains perturbatively stable and therefore continues to provide a meaningful phenomenological constraint.

The importance of positivity has become increasingly evident in recent NNLO polarized PDF determinations. For example, the NNPDFpol2.0 analysis~\cite{Cruz-Martinez:2025ahf} showed that positivity constraints significantly reduce uncertainties in poorly constrained regions, particularly for the polarized gluon distribution. Similar observations have been reported in JAM analyses~\cite{Sato:2016tuz}, where positivity helps control the large-\(x\) behavior of helicity PDFs and improves the overall stability of the fit.

Although recent NNLO analyses include a wide range of observables such as SIDIS and polarized proton--proton scattering~\cite{Bertone:2024taw,Cruz-Martinez:2025ahf}, inclusive polarized DIS remains the theoretically cleanest process. Restricting the present study to inclusive DIS therefore provides a transparent framework for studying NNLO QCD effects and a comparison with our previous NLO analyses~\cite{Khorramian:2020gkr}.

Motivated by the increasing precision of polarized scattering data and future measurements at the Electron--Ion Collider (EIC)~\cite{Accardi:2012qut}, NNLO accuracy is essential for reducing theoretical uncertainties and improving the reliability of helicity PDF extractions.

In this work, we present an NNLO QCD analysis of polarized parton distribution functions based exclusively on inclusive polarized DIS data. The analysis extends the KLSS21 NLO framework~\cite{Khorramian:2020gkr} to NNLO accuracy using the \texttt{APFEL++}~\cite{Bertone:2017gds}  evolution package, supplemented by our own consistent treatment of target-mass corrections, higher-twist effects, and positivity constraints. A central objective of the present study is to obtain improved constraints on the polarized strange-quark distribution and to investigate the sensitivity of the extracted helicity PDFs to both the higher-twist parametrization and the axial charge assumptions. To this end, we compare additive and multiplicative higher-twist corrections and examine the stability of the results under variations of the nonsinglet axial charges \(a_3\) and \(a_8\), thereby assessing the impact of possible SU(3)-flavor symmetry breaking.

The remainder of this paper is organized as follows. In
Sec.~\ref{sec:Formalism}, we present the NNLO formalism for polarized
deep-inelastic scattering, including the general
theoretical framework and the phenomenological treatment
of higher-twist effects in both additive and
multiplicative forms. Section~\ref{sec:Parametrization} introduces the
parametrization of the polarized PDFs at the initial
scale, together with the implementation of the nonsinglet
axial charge constraints, positivity conditions, and the
numerical setup of the fit. In Sec.~\ref{sec:data}, we describe the
polarized DIS data sets included in the analysis, as well
as the corresponding kinematic coverage and selection
criteria. The results of the NNLO analysis are presented
in Sec.~\ref{sec:results}, where we discuss the extracted polarized PDFs,
the higher-twist contributions, and the cross-checks
based on fits with free axial charges. Finally, the
availability of the polarized PDF sets and our concluding
remarks are given in the last section.

\section{NNLO Formalism for Polarized Structure Functions} \label{sec:Formalism}

In this section, we present the theoretical framework employed to determine polarized parton distribution functions at NNLO accuracy. Particular emphasis is placed on the treatment of  HT effects, which are essential for a reliable description of polarized DIS data in the preasymptotic region. We also introduce the additive and multiplicative higher-twist parametrizations used throughout the analysis to investigate the sensitivity of the extracted helicity PDFs, especially the polarized strange quark distribution, to the modeling of power-suppressed contributions.

\subsection{Polarized Structure Functions at NNLO}

A characteristic feature of polarized DIS is that a significant fraction of the available experimental data lies in the moderate-$Q^2$ and preasymptotic region, typically
$
Q^2 \sim 1\text{--}5~\mathrm{GeV}^2
$
and $4~\mathrm{GeV}^2 \lesssim W^2 \lesssim 10~\mathrm{GeV}^2$, where $W^2=M^2+Q^2\left(\frac{1}{x}-1\right)$ is the invariant mass squared of the hadronic final state, and $M$ is the nucleon mass.
This kinematic region is particularly relevant for high-precision Jefferson Lab measurements included in the present analysis. Due to the limited amount of polarized data compared to the unpolarized case, such regions cannot be excluded through restrictive kinematic cuts. Therefore, a consistent theoretical description of the polarized structure function $g_1(x,Q^2)$ requires the inclusion of both target-mass and higher-twist contributions~\cite{Leader:2005ci,Khorramian:2020gkr}.

Within perturbative QCD, the polarized structure function can be decomposed as
\begin{equation}
g_1(x,Q^2)=
g_1^{\mathrm{LT}}(x,Q^2)
+
g_1^{\mathrm{HT}}(x,Q^2)\,,
\label{eq:g1_decomposition}
\end{equation}
where $g_1^{\mathrm{LT}}$ denotes the leading-twist contribution, while $g_1^{\mathrm{HT}}$ encodes dynamical higher-twist power corrections arising from operators of twist $\tau \ge 3$.

The leading-twist contribution at NNLO accuracy is written as
\begin{eqnarray}
g_1^{\mathrm{LT}}(x,Q^2)
&=&
g_1^{\rm pQCD}(x,Q^2)
+
\frac{h^{\rm TMC}(x,Q^2)}{Q^2}
\nonumber\\
&&+
{\cal O}(M^4/Q^4)\,,
\label{g1LT}
\end{eqnarray}
where the target-mass corrections $h^{\rm TMC}(x,Q^2)$ are purely kinematic effects and are exactly calculable within QCD~\cite{TMC}. In this work, TMC effects are consistently included in the leading-twist term, while $g_1^{\mathrm{HT}}$ accounts exclusively for the remaining dynamical power corrections beyond TMC.

The leading-twist perturbative contribution to the polarized structure function is calculated at NNLO accuracy. In the \(\overline{\rm MS}\) factorization scheme, \(g_1^{\rm pQCD}(x,Q^2)\) can be written as
\begin{eqnarray}
g_1^{\rm pQCD}(x,Q^2)
&=&\
\frac{1}{2}
\sum_{q=1}^{n_f} e_q^2
\Bigg[
(\Delta q+\Delta\bar q)
\otimes
\Bigg(
1+\frac{\alpha_s}{2\pi}\Delta C_q^{(1)}
\nonumber\\
&+&
\frac{\alpha_s^2}{4\pi^2}\Delta C_q^{(2)}
\Bigg)
+
\Delta g \otimes
\Bigg(
\frac{\alpha_s}{2\pi}\Delta C_g^{(1)}
\nonumber\\
&+&
\frac{\alpha_s^2}{4\pi^2}\Delta C_g^{(2)}
\Bigg)
\Bigg]\,,
\label{eq:g1_pqcd_NNLO}
\end{eqnarray}
where \(e_q\) denotes the electric charge of the quark flavor \(q\), while
\(\Delta q(x,Q^2)\),
\(\Delta \bar q(x,Q^2)\),
and
\(\Delta g(x,Q^2)\)
represent the polarized quark, antiquark, and gluon distributions, respectively. The polarized coefficient functions
\(\Delta C_i^{(1)}\)
and
\(\Delta C_i^{(2)}\)
correspond to the NLO and NNLO spin-dependent Wilson coefficient functions computed in the
\(\overline{\mathrm{MS}}\)
factorization scheme~\cite{Mertig:1995ny,Vogelsang:1995vh,Zijlstra:1993sh}. The NNLO coefficient functions
\(\Delta C_i^{(2)}\)
are fully included in the present analysis, and the corresponding polarized splitting functions governing the DGLAP evolution equations are consistently implemented at NNLO accuracy~\cite{Moch:2014sna,Vogt:2008yw}.

The symbol \(\otimes\) denotes the standard Mellin convolution in Bjorken-\(x\) space,
\begin{equation}
[f \otimes g](x)
=
\int_x^1
\frac{dy}{y}\,
f(y)\,
g\!\left(\frac{x}{y}\right).
\label{eq:mellin_conv}
\end{equation}

\subsection{Phenomenological Treatment of Higher-Twist Effects}

In the operator product expansion (OPE) approach, contributions beyond the leading-twist approximation emerge as power-suppressed corrections to the DIS structure functions.  These terms originate from multi-parton correlations and other nonperturbative QCD dynamics and become increasingly important at moderate values of $Q^2$. Because a substantial fraction of the available polarized DIS data is collected in this kinematic region, higher-twist contributions must be taken into account alongside the leading-twist term in any realistic QCD analysis.

While the perturbative contribution \(g_1^{\rm pQCD}(x,Q^2)\) is fully known up to NNLO accuracy, the treatment of higher-twist power corrections is necessarily phenomenological. Different implementations of higher-twist effects have been proposed in the literature, reflecting the fact that the corresponding nonperturbative matrix elements cannot presently be calculated with sufficient precision from first principles. A central objective of the present work is therefore to assess the sensitivity of the extracted helicity PDFs to the phenomenological modeling of higher-twist effects. To this end, two independent realizations of higher-twist corrections are considered and systematically compared throughout the analysis. Unlike the KLSS21 NLO analysis~\cite{Khorramian:2020gkr}, where only the additive parametrization was considered, the present NNLO analysis examines both additive and multiplicative parametrizations to quantify the residual theoretical uncertainty associated with the treatment of power-suppressed contributions.

In the conventional additive formulation, the higher-twist
contribution is parametrized as
\begin{equation}
g_1^{\mathrm{add}}(x,Q^2)
=
g_1^{\mathrm{LT}}(x,Q^2)
+
\frac{H(x)}{Q^2}\,,
\label{eq:g1_HT_add}
\end{equation}
where $H(x)$ denotes an effective higher-twist function
associated with nonperturbative power-suppressed
corrections. In this parametrization, the higher-twist term
is assumed to be independent of the leading-twist structure
function and does not introduce any explicit $Q^2$
dependence beyond the overall $1/Q^2$ suppression.

Alternatively, higher-twist effects can be incorporated using a multiplicative parametrization,
\begin{eqnarray}
g_1^{\mathrm{mult}}(x,Q^2)
&=&
g_1^{\mathrm{LT}}(x,Q^2)
\left(
1+\frac{C(x)}{Q^2}
\right),
\label{eq:g1_HT_mult}
\end{eqnarray}
where $C(x)$ represents an effective higher-twist
correction factor.

It is important to emphasize that the additive and
multiplicative implementations correspond to different
assumptions regarding the $Q^2$ behavior of higher-twist
effects. Indeed, the multiplicative expression of
Eq.~(\ref{eq:g1_HT_mult}) can be rewritten in an additive
form,
\begin{equation}
g_1^{\mathrm{mult}}(x,Q^2)
=
g_1^{\mathrm{LT}}(x,Q^2)
+
\frac{\widetilde H(x,Q^2)}{Q^2}\,,
\label{eq:g1_HT_equiv}
\end{equation}
with
\begin{equation}
\widetilde H(x,Q^2)
=
g_1^{\mathrm{LT}}(x,Q^2)\,C(x)\,.
\label{eq:Htilde}
\end{equation}

Unlike the additive higher-twist function $H(x)$ in
Eq.~(\ref{eq:g1_HT_add}), the effective coefficient
$\widetilde H(x,Q^2)$ inherits both the scale dependence
and flavor structure of the leading-twist polarized
structure function. Consequently, the additive and
multiplicative parametrizations should not be regarded as
equivalent representations of the same higher-twist
function. Rather, they correspond to different realizations
of power corrections and may lead to different extracted
higher-twist coefficients, even when the resulting
polarized PDFs remain compatible within uncertainties.

For this reason, we perform independent analyses using
both higher-twist implementations and use their
comparison as a measure of the sensitivity of the
extracted results to the phenomenological modeling of
higher-twist effects.

For simplicity, no explicit logarithmic \(Q^2\)-dependence is assigned to the effective higher-twist functions. This approximation is motivated by the expectation that the dominant behavior is governed by the overall \(1/Q^2\) suppression. At the same time, any residual logarithmic dependence is likely to remain within the current experimental uncertainties. Consequently, the extracted quantities \(H(x_i)\) and \(C(x_i)\) should be interpreted as effective higher-twist coefficients corresponding to the characteristic \(Q^{2}\) values associated with the individual \(x_i\) bins~\cite{Leader:2005ci,Khorramian:2020gkr}.

Generally, two main approaches are considered in the literature to parameterize the phenomenological higher-twist coefficient functions. In the first approach, a flexible functional form (such as a polynomial) is adopted over the entire range of $x$, where higher powers of $x$ effectively control the $x \to 1$ region~\cite{Accardi:2023gyr, Shahrzad:2024ucw, Accardi:2026hdv}. This approach is primarily used in unpolarized analyses, where the abundant data can tightly constrain the large-$x$ behavior. The second approach, widely employed in both polarized and unpolarized analyses, parameterizes the HT functions using spline polynomials or linear combinations defined over discrete $x$ nodes, with the values at each node treated as free fit parameters~\cite{Khorramian:2020gkr, Alekhin:2022uwc, Ball:2025xtj}. This flexible method localized the $x$-dependence of the HT corrections, making it particularly advantageous for polarized analyses where data in the large-$x$ region remain experimentally sparse. In this work, we adopt the second approach and parameterize both the additive and multiplicative HT functions using linear interpolation across the node grid $x_i \in \{0.003, 0.05, 0.15, 0.25, 0.40, 0.75\}$.

Several insightful studies in the {\it unpolarized} sector have investigated the impact of HT corrections. For instance, Ref.~\cite{Cerutti:2025yji} examines the interplay between isospin-dependent effects and different implementations of HT corrections (additive versus multiplicative). In Refs.~\cite{Harland-Lang:2025wvm,Ball:2025xtj}, the authors investigate missing higher-order uncertainties (MHOUs), which are closely related to the treatment of HT contributions in phenomenological analyses. In particular, Ref.~\cite{Ball:2025xtj} proposes an approach in which HT contributions are incorporated directly at the level of observables.

\section{Polarized PDF Parametrization and Fit Setup} \label{sec:Parametrization}

In this section, we describe the parametrization of the polarized parton distribution functions at the input scale and the theoretical constraints imposed in the analysis. Particular attention is given to the implementation of axial-vector sum rules, possible SU(3) flavor symmetry breaking effects, and positivity constraints. We also summarize the numerical setup adopted for the NNLO QCD analysis and the determination of PDF uncertainties.

\subsection{PDF Parametrization}

Following the KLSS21 NLO analysis~\cite{Khorramian:2020gkr}, we take $Q_0^2 = 1~\mathrm{GeV}^2$ for  the quark-plus-antiquark helicity combinations,
\begin{equation}
\Delta q^{+}(x,Q^2)=\Delta q(x,Q^2)+\Delta \bar q(x,Q^2)\,,
\qquad q=u,d,s.
\end{equation}
The same functional form is retained in the present NNLO study to allow a direct assessment of the impact of higher-order perturbative corrections, higher-twist modeling, and possible SU(3) flavor symmetry-breaking effects on the extracted polarized PDFs.

At the input scale, the polarized PDFs are parametrized as
\begin{align}
x\Delta u^{+}(x,Q_0^2)
&=
A_{u_{+}}
x^{\alpha_{u_{+}}}
(1-x)^{\beta_{u_{+}}}
\left(
1+\epsilon_{u_{+}}\sqrt{x}
+\gamma_{u_{+}}x
\right),
\nonumber\\
x\Delta d^{+}(x,Q_0^2)
&=
A_{d_{+}}
x^{\alpha_{d_{+}}}
(1-x)^{\beta_{d_{+}}}
\left(
1+\epsilon_{d_{+}}\sqrt{x}
+\gamma_{d_{+}}x
\right),
\nonumber\\
x\Delta s^{+}(x,Q_0^2)
&=
A_{s_{+}}
x^{\alpha_{s_{+}}}
(1-x)^{\beta_{s_{+}}}
\left(
1+\gamma_{s_{+}}x
\right),
\nonumber\\
x\Delta g(x,Q_0^2)
&=
A_{g}
x^{\alpha_{g}}
(1-x)^{\beta_{g}}
\left(
1+\gamma_{g}x
\right).
\label{input_PDFs}
\end{align}

As in the KLSS21 analysis~\cite{Khorramian:2020gkr}, no additional assumptions are imposed on the light sea-quark polarizations, \(\Delta\bar u\) and \(\Delta\bar d\). The fit therefore determines the helicity combinations \(\Delta q^{+}\) directly from polarized DIS data.

In addition to the normalization parameters and the small-
and large-$x$ exponents, the input parametrization
contains extra shape parameters that control the
flexibility of the polarized distributions. In practice,
the current inclusive polarized DIS data can
constrain only a limited subset of these shape
coefficients. As a result, the full functional form is
not used for all parton flavors.

For the light-quark combinations $\Delta u^{+}$ and
$\Delta d^{+}$, the inclusion of the $\sqrt{x}$ term
provides additional flexibility and improves the stability
of the fit and the determination of the corresponding
parameter uncertainties. In contrast, for the polarized
strange-quark and gluon distributions, the present data
do not support a statistically meaningful determination
of the corresponding $\sqrt{x}$ coefficients. We
therefore omit the $\epsilon$ terms in the
parametrizations of $\Delta s^{+}$ and $\Delta g$. 

Similarly, the available polarized DIS data do not
justify introducing more elaborate functional forms for
all partonic species. The final parametrization adopted
in Eq.~\eqref{input_PDFs} thus represents a compromise
between flexibility and fit stability, retaining only
those shape parameters that are supported by the data and
avoiding the introduction of poorly constrained degrees
of freedom.

\subsection{Axial Charges and SU(3) Constraints}
The flavor decomposition of polarized parton distribution
functions is constrained by fixing the values of the nonsinglet axial-vector
charges associated with QCD flavor symmetries. These
constraints are implemented through the axial charge sum
rules
\begin{equation}
a_3 =
\int_0^1 dx \,
\left[
\Delta u^{+}(x,Q^2)-\Delta d^{+}(x,Q^2)
\right],
\label{eq:a3moment}
\end{equation}
and
\begin{equation}
a_8 =
\int_0^1 dx \,
\left[
\Delta u^{+}(x,Q^2)+\Delta d^{+}(x,Q^2)-2\Delta s^{+}(x,Q^2)
\right],
\label{eq:a8moment}
\end{equation}
where \(a_3\) and \(a_8\) denote the triplet and octet
axial charges, respectively, whose values, as explained below, are supposed to be relatively well determined.

In the standard polarized-PDF analysis, these sum rules
are imposed using phenomenological  values for $a_3$ and $a_8$ extracted from
low-energy weak decays. The triplet axial charge is
precisely determined from neutron \(\beta\)-decay \cite{PDG},
\begin{equation}
a_3 = g_A = F + D = 1.2756 \pm 0.0013\,,
\label{eq:a3value}
\end{equation}
and therefore provides a strong constraint on the
flavor-nonsinglet combination
\(\Delta u^{+}-\Delta d^{+}\).
The octet axial charge is conventionally taken as 
\begin{equation}
a_8 = 3F-D = 0.585 \pm 0.025\,,
\label{eq:a8value}
\end{equation}
as extracted from hyperon semileptonic decays under the
assumption of approximate SU(3)-flavor symmetry \cite{PDG}.

Unlike \(a_3\), the octet axial charge is potentially more
sensitive to SU(3)-flavor symmetry breaking. This is
particularly important for polarized PDF analyses because
\(a_8\) directly constrains the combination
\(\Delta u^{+}+\Delta d^{+}-2\Delta s^{+}\), and hence has
a direct impact on the determination of the polarized
strange-quark distribution \(\Delta s^{+}\). As a result,
the strange-quark helicity density is expected to be the
parton distribution most sensitive to variations in the
assumed value of \(a_8\).

As an additional cross-check of the stability of
the fit, we also perform an alternative analysis in which
the two axial charges \(a_3\) and \(a_8\) are treated as free
fit parameters and determined directly from the polarized
DIS data.

The impact of this alternative fit on the extracted
polarized PDFs, and in particular on the polarized
strange-quark distribution, will be examined in Sec.\ref{subsec:free_axial_charges} as
a dedicated stability test of the axial charge
assumptions.

\subsection{Positivity Constraints}

Positivity provides an important phenomenological
constraint on polarized parton distributions,
particularly in kinematic regions where the experimental
information is limited. In the present analysis, we
impose the leading-order positivity condition given in
Eq.~(\ref{eq:positivity}), using the corresponding
unpolarized PDFs as upper bounds on the polarized parton
densities.

Since our analysis is performed in terms of
quark-plus-antiquark combinations,
$\Delta q^+=\Delta q
+\Delta\bar q$, where $q=u,d,s$, rather than
separate quark and antiquark distributions, the positivity
constraint is applied in the form
$|\Delta q^+|<q^+$, where
$q^+=q+\bar q$.
For the gluon distribution, the analogous condition
$|\Delta g|<g$ is imposed. This constraint is especially relevant for the
strange-quark and gluon distributions, which remain less
directly constrained by the presently available
polarized inclusive DIS data.

Beyond leading order, PDFs become factorization-scheme dependent, and the positivity condition strictly applies to physical observables rather than to the PDFs themselves. Nevertheless, positivity constraints remain a valuable phenomenological tool and are widely used in polarized PDF analyses to stabilize the extraction of poorly constrained distributions, particularly the polarized strange-quark and gluon densities.

In the kinematic region relevant to the present analysis,
this phenomenological implementation provides a stable and
practically useful constraint on the extracted polarized
PDFs, especially in sectors that are only weakly
constrained by inclusive DIS data.

In the present NNLO analysis, positivity constraints are implemented using the MSHT20 NNLO unpolarized parton distributions~\cite{Bailey:2020ooq}. Compared to the KLSS21 NLO analysis, which employed the MMHT14 NLO set~\cite{MMHT14}, this choice ensures consistency between the perturbative order of the polarized fit and that of the unpolarized reference distributions used in the positivity conditions.

In practice, to facilitate the implementation of the positivity constraints for the polarized strange-quark and gluon distributions, we fix
\(
\beta_{s^{+}} = 17.99, 
\beta_g = 2.1278.
\)
These values are guided by the large-$x$ behavior of the MSHT20 NNLO unpolarized PDFs and are kept fixed throughout the fit. This choice stabilizes the large-$x$ behavior of the polarized strange-quark and gluon distributions and ensures the fulfillment of the positivity constraints over the kinematic region covered by the data.

\subsection{Fit Strategy and Numerical Setup}

In polarized DIS, the kinematic coverage and precision of the available data remain significantly more limited than in the unpolarized case. Therefore, the strong coupling constant is not treated as a free fit parameter but is taken from the MSHT20 NNLO analysis. In particular, we adopt
\(
\alpha_s(M_Z^2)=0.118,
\)
and evolve the coupling consistently at NNLO accuracy. This choice ensures compatibility between the evolution of the polarized PDFs and the unpolarized distributions entering the positivity constraints, thereby providing a fully consistent NNLO framework for the determination of polarized parton densities.

The numerical implementation of the NNLO evolution equations and polarized structure functions is performed using the \texttt{APFEL++} framework~\cite{Bertone:2017gds}. This program provides an accurate and efficient solution of the DGLAP evolution equations and allows for a fully consistent treatment of polarized DIS observables at NNLO accuracy. For consistency with the MSHT20 NNLO~\cite{Bailey:2020ooq} unpolarized PDFs used
in the positivity constraints, we adopt the charm and
bottom quark masses from the MSHT20 framework,
\(
m_c = 1.40~\mathrm{GeV},
m_b = 4.75~\mathrm{GeV}
\).
The minimization of the global \(\chi^2\) function is performed using the CERN MINUIT package \cite{James:1975dr}. PDF uncertainties are estimated using the standard Hessian approach with a tolerance corresponding to
\(
\Delta \chi^2 = 1.
\)

\section{Polarized DIS Data Sets} \label{sec:data}

In this section, we summarize the polarized inclusive DIS data sets included in the analysis and discuss the corresponding kinematic coverage. We also describe the kinematic selections adopted in the fit and the role of the Jefferson Lab measurements in constraining polarized PDFs in the preasymptotic region.


The present analysis is based primarily on world data for polarized inclusive DIS, which provide the most direct experimental constraints on helicity-dependent PDFs. The data set includes measurements of polarized structure functions obtained from several fixed-target experiments performed at CERN, SLAC, DESY, and Jefferson Lab.

In particular, we consider polarized DIS measurements from the EMC~\cite{EMC}, SMC~\cite{SMC}, and COMPASS~\cite{COMPASS1,COMPASS2} collaborations at CERN, from the E142~\cite{E142}, E143~\cite{E143}, E154~\cite{E154}, and E155~\cite{E155} experiments at SLAC, from the HERMES~\cite{HERMES1} experiment at DESY, and from the Hall A~\cite{HallA} and CLAS~\cite{CLAS} collaborations at Jefferson Lab. These experiments provide measurements of the polarized structure function \(g_1(x,Q^2)\) for proton, deuteron, and neutron targets over a broad kinematic range in \(x\) and \(Q^2\).

A summary of the experimental data sets included in the present fit is presented in Table~\ref{tab:table1}. The selected data points cover approximately the kinematic region $0.004 \lesssim x \lesssim 0.8$, and $1 \lesssim Q^2 \lesssim 100~\mathrm{GeV}^2$.

\begin{table*}[ht]
\caption{\label{tab:table1}
Comparison of the partial $\chi^2$ values obtained using additive and multiplicative higher-twist corrections in the NNLO QCD analysis of polarized inclusive DIS data.
}
\begin{ruledtabular}
\begin{tabular}{lcccc}
\multirow{2}{*}{Experiment}
&
\multirow{2}{*}{Process}
&
\multirow{2}{*}{$N_{\rm data}$}
&
\multicolumn{2}{c}{$\chi^2$}
\\
\cline{4-5}
&&&
Additive HT
&
Multiplicative HT
\\
\hline

EMC~\cite{EMC}              & DIS(p) & 10  & 7.02  & 7.39  \\
SMC~\cite{SMC}              & DIS(p) & 12  & 4.74  & 5.44  \\
SLAC/E143~\cite{E143}       & DIS(p) & 28  & 24.09 & 24.72 \\
SLAC/E155~\cite{E155}       & DIS(p) & 24  & 20.39 & 18.91 \\
HERMES~\cite{HERMES1}       & DIS(p) & 37  & 18.21 & 18.44 \\
COMPASS'10~\cite{COMPASS1}  & DIS(p) & 15  & 11.58 & 11.99 \\
COMPASS'16~\cite{COMPASS2}  & DIS(p) & 51  & 30.20 & 30.40 \\
CLAS/EG1b~\cite{CLAS}       & DIS(p) & 166 & 93.49 & 93.86 \\
\hline

SMC~\cite{SMC}              & DIS(d) & 12  & 17.51 & 18.33 \\
SLAC/E143~\cite{E143}       & DIS(d) & 28  & 41.30 & 40.58 \\
SLAC/E155~\cite{E155}       & DIS(d) & 24  & 18.49 & 18.89 \\
HERMES~\cite{HERMES1}       & DIS(d) & 37  & 36.88 & 35.53 \\
COMPASS'17~\cite{COMPASS2}  & DIS(d) & 43  & 28.82 & 30.20 \\
CLAS/EG1b~\cite{CLAS}       & DIS(d) & 158 & 134.19 & 129.46 \\
\hline

SLAC/E142~\cite{E142}       & DIS(n) & 8   & 6.29 & 4.88 \\
SLAC/E154~\cite{E154}       & DIS(n) & 17  & 6.04 & 8.41 \\
HERMES~\cite{HERMES1}       & DIS(n) & 9   & 2.69 & 2.78 \\
JLab Hall A~\cite{HallA}    & DIS(n) & 3   & 2.47 & 1.57 \\
\hline\hline

\textbf{Total}
&
&
\text{682}
&
\text{504.41}
&
\text{501.77}
\\
\hline

\textbf{$\chi^2/\mathrm{n.d.o.f.}$}
&
&
\textbf{}
&
\text{0.764}
&
\text{0.760}

\end{tabular}
\end{ruledtabular}
\end{table*}

\begin{table*}[t]
\caption{\label{tab:NNLOparams}
Parameters of the NNLO fitted polarized PDFs at
$Q_0^2=1~{\rm GeV}^2$
obtained using additive and multiplicative higher-twist
approaches.
 The fits are performed using the fixed axial charge
values $a_3$ and $a_8$ defined in
Eqs.~(\ref{eq:a3value}) and (\ref{eq:a8value}).
Parameters marked by (*) are fixed.
}
\begin{ruledtabular}
\begin{tabular}{lccccc}
Flavor &
$A$ &
$\alpha$ &
$\beta$ &
$\epsilon$ &
$\gamma$
\\
\hline

\multicolumn{6}{c}{Additive HT} \\
\hline
$u^{+}$
& $1.7494$
& $0.7378\pm0.1070$
& $3.2670\pm0.1898$
& $-3.0354\pm0.4421$
& $8.1265\pm1.7745$ \\

$d^{+}$
& $-5.8164$
& $1.1902\pm0.2107$
& $5.3608\pm1.1278$
& $-2.5491\pm0.8489$
& $3.6564\pm1.6360$ \\

$s^{+}$
& $-2.4003\pm2.6736$
& $1.4238\pm0.3564$
& $17.9930^{*}$
& --
& -- \\

$g$
& $1.9325\pm1.9170$
& $1.8649\pm0.7248$
& $2.1278^{*}$
& --
& -- \\

\hline
\multicolumn{6}{c}{Multiplicative HT} \\
\hline
$u^{+}$
& $1.6742$
& $0.6884\pm0.0952$
& $3.4275\pm0.1908$
& $-3.5501\pm0.4027$
& $9.3278\pm1.8725$ \\

$d^{+}$
& $-7.3779$
& $1.1948\pm0.3220$
& $7.1713\pm1.5660$
& $-3.7480\pm1.1392$
& $6.9476\pm2.0628$ \\

$s^{+}$
& $-0.9401\pm3.2149$
& $1.3769\pm1.0856$
& $17.9930^{*}$
& --
& -- \\

$g$
& $2.2017\pm2.3624$
& $2.1788\pm0.8699$
& $2.1278^{*}$
& --
& -- \\
\end{tabular}
\end{ruledtabular}
\end{table*}

\begin{figure}[!htbp]
    \centering
    \includegraphics[width=0.95\linewidth,clip]{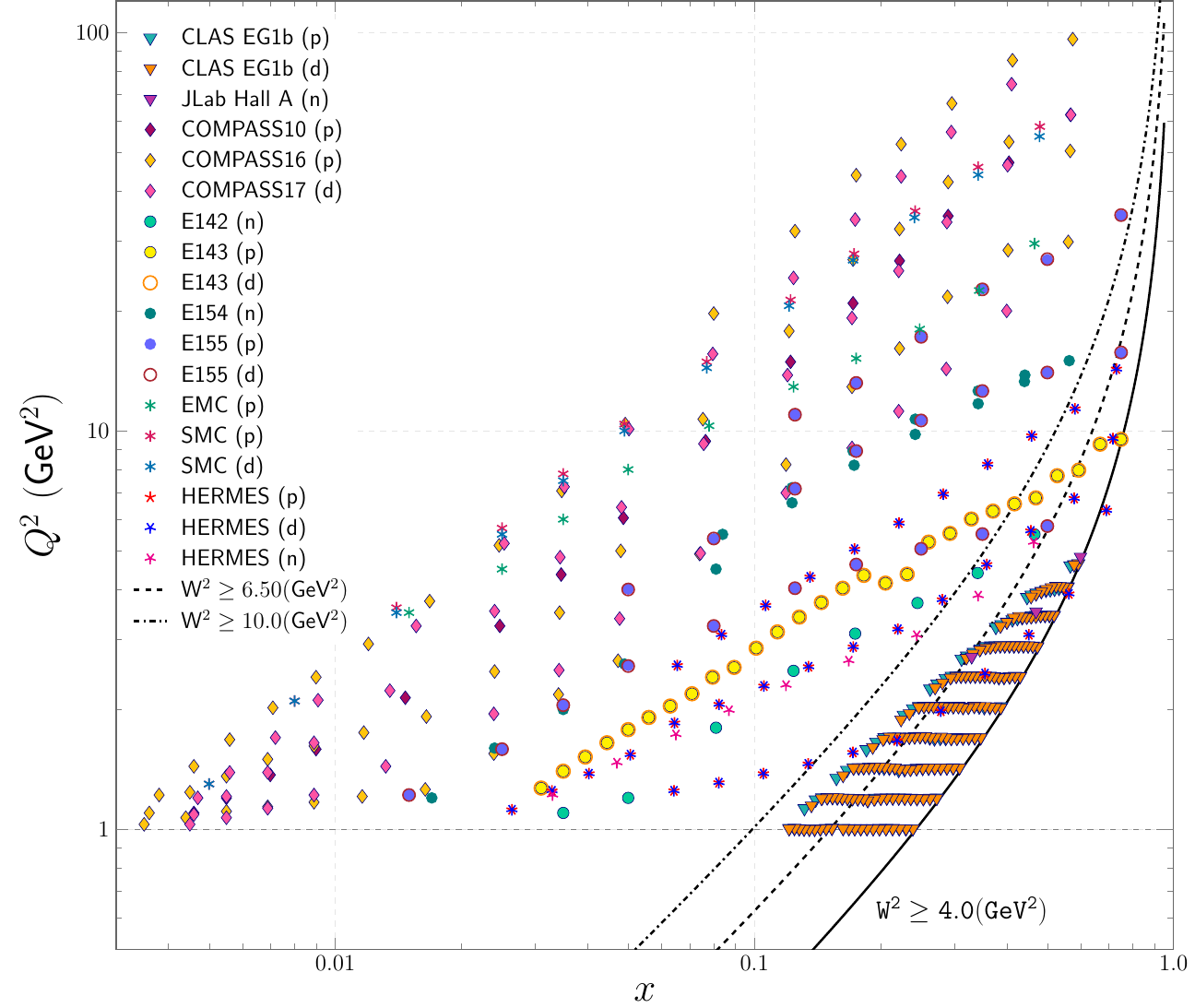}
	\caption{Kinematic coverage in the $(x, Q^2)$ plane for the datasets included in this analysis. The horizontal dashed line denotes the baseline momentum transfer cut of $Q_0^2 = 1.0~\mathrm{GeV}^2$. The solid line indicates the relaxed invariant mass cut $W^2 \geq 4.0~\mathrm{GeV}^2$, adopted here to incorporate additional large-$x$ data. For comparison, the stricter constraints $W^2 \geq 6.5~\mathrm{GeV}^2$ and $W^2 \geq 10.0~\mathrm{GeV}^2$ utilized by other global analyses are also shown.}
	\label{fig:xq2_coverage}
\end{figure}

A characteristic feature of the presently available polarized DIS measurements is that a substantial fraction of the data lies in the moderate-\(Q^2\) and preasymptotic region, particularly in the case of the high-precision Jefferson Lab measurements. In contrast to unpolarized global analyses, where such kinematic regions are frequently removed through restrictive cuts, polarized DIS studies cannot simply exclude these data because of the comparatively limited amount of precise polarized measurements. Consequently, TMC and HT contributions must be included in order to achieve a reliable description of the experimental observables and a stable extraction of helicity PDFs, especially in the moderate- and large-\(x\) regions.

To ensure the applicability of perturbative QCD and the validity of our higher-twist framework, we impose kinematic cuts on the virtuality $Q^2$ and the hadronic invariant-mass squared variable $W^2$.  Specifically, we restrict the data to $Q^2 >Q^2_{cut} = 1~\mathrm{GeV}^2$ and $W^2 >W^2_{cut} =4~\mathrm{GeV}^2$. Here, we emphasize the distinct features of this analysis in comparison to the recent MAPPDFpol1.0 analysis~\cite{Bertone:2024taw}, where a stricter cut of $W^2 > 6.5~\mathrm{GeV}^2$ was applied, and the Hall A~\cite{HallA} and CLAS~\cite{CLAS} data are completely excluded from the fit.

These kinematic selections are particularly important for the consistent treatment of TMC and HT effects within the NNLO framework adopted in this work. The importance of the Jefferson Lab measurements for constraining polarized PDFs and studying the sensitivity of the strange quark polarization to flavor SU(3) symmetry breaking was already emphasized in the previous KLSS21 analysis~\cite{Khorramian:2020gkr}.

It should be noted that NNLO polarized PDF analyses based on more restrictive kinematic selections may exclude a significant fraction of the precise Jefferson Lab Hall A~\cite{HallA} and CLAS~\cite{CLAS} polarized DIS measurements~\cite{Bertone:2024taw}. In the present work, however, these measurements are retained in the fit since they provide important sensitivity to TMC and HT effects in the preasymptotic region. This is particularly important for studying the interplay between NNLO QCD corrections and nonperturbative HT contributions in polarized proton and neutron structure functions.

As already observed in the KLSS21 analysis~\cite{Khorramian:2020gkr}, the Jefferson Lab CLAS/EG1b proton and deuteron measurements remain compatible with the global polarized DIS data set and continue to provide important constraints on helicity PDFs in the preasymptotic region.

The kinematic range spanned by the experimental data sets included in our analysis is shown in Fig.~\ref{fig:xq2_coverage}. As already mentioned, to ensure the applicability of perturbative QCD, we adopt a minimum input scale of
$
Q_0^2 = 1.0~\mathrm{GeV}^2.
$
While many global analyses impose stringent kinematic cuts, such as
$W^2 \geq 6.5~\mathrm{GeV}^2$ or even
$W^2 \geq 10.0~\mathrm{GeV}^2$, to suppress higher-twist and target-mass effects, we instead employ a more relaxed requirement of
$W^2 \geq 4.0~\mathrm{GeV}^2$. As illustrated in Fig.~\ref{fig:xq2_coverage}, this choice substantially increases the number of retained data points in the large-$x$ region. Preserving these measurements is essential for constraining the polarized PDFs as $x \rightarrow 1$, allowing our analysis to probe the large-$x$ behavior that is often excluded by more restrictive kinematic selections.

\section{Results and Discussion} \label{sec:results}

In this section, we present the main results of the NNLO analysis of polarized parton distribution functions and discuss the impact of different HT  implementations. The additive HT fit is taken as the baseline determination, and its results are also presented in comparison with the  KLSS21~(NLO) results, while the multiplicative form is used to assess the residual model dependence associated with power-suppressed corrections. 

The quality of the global fit is summarized in
Table~\ref{tab:table1}, where the partial and total
$\chi^2$ values are shown for both higher-twist
parametrizations. Both approaches provide a good overall description of the
world polarized inclusive DIS data and yield very similar
fit qualities across all experimental data sets included
in the analysis.

The total $\chi^2$ decreases from
\(
\chi^2 = 504.41
\)
for the additive fit to
\(
\chi^2 = 501.77
\)
for the multiplicative fit, corresponding to a difference
of less than three units for 682 data points. Such a small
variation indicates that the present polarized DIS data do
not exhibit a strong preference for either higher-twist
implementation.

A detailed comparison of the partial $\chi^2$
contributions indicates that both additive and
multiplicative higher-twist parametrizations describe the
proton, deuteron, and neutron data with comparable
quality.

The corresponding input PDF parameters are listed in
Table~\ref{tab:NNLOparams}. Although some individual fit
parameters differ between the additive and multiplicative
solutions, the resulting parametrizations lead to
comparable descriptions of the experimental data. This
indicates that the observed parameter variations largely
reflect correlations among the fitted coefficients rather
than genuine physical differences in the extracted
polarized parton distributions.

\begin{figure}[t]
\centering
\includegraphics[width=0.38\textwidth]{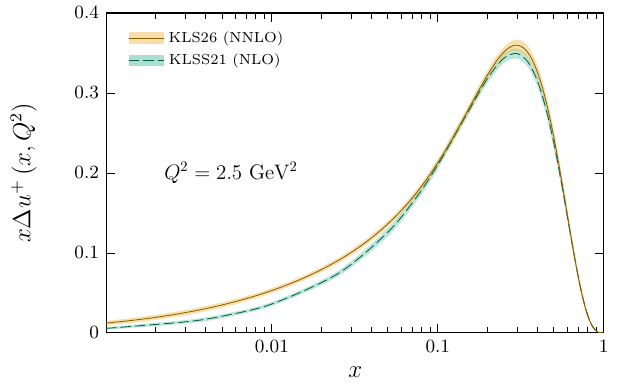}
\includegraphics[width=0.38\textwidth]{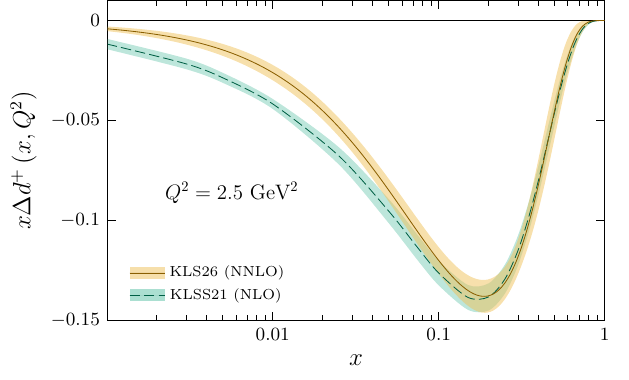}
\\
\includegraphics[width=0.38\textwidth]{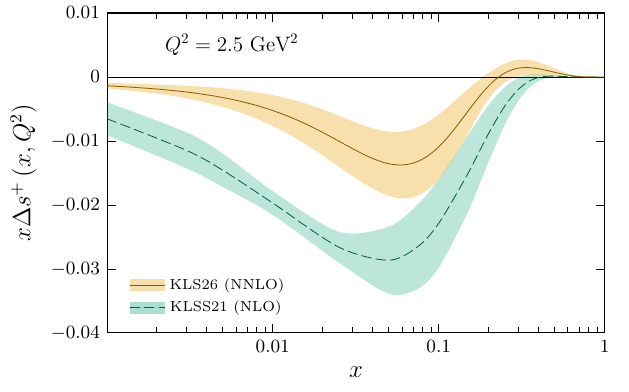}
\includegraphics[width=0.38\textwidth]{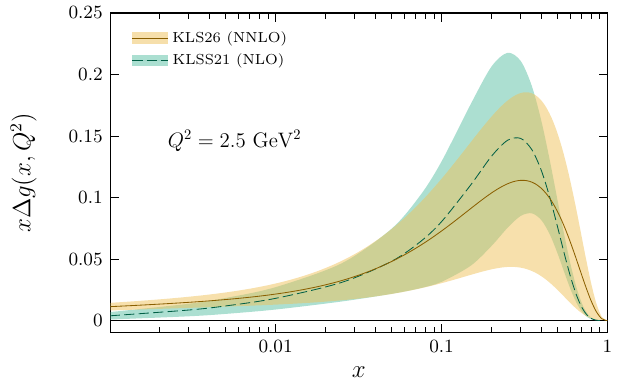}
\caption{
Polarized parton distribution functions obtained in the
present NNLO analysis at $Q^2=2.5~{\rm GeV}^2$ using the
additive higher-twist parametrization,
compared with the previous KLSS21 NLO determination~\cite{Khorramian:2020gkr}. The panels show $x\Delta q^{+}=x(\Delta q+\Delta\bar q)$
for $q=u,d,s$ and $x\Delta g$. The uncertainty bands correspond to the PDF
uncertainties obtained from the Hessian analysis.
}
\label{fig:NLO_NNLO_PDFs}
\end{figure}

\begin{figure*}[htbp]
\centering

\includegraphics[width=0.38\textwidth]{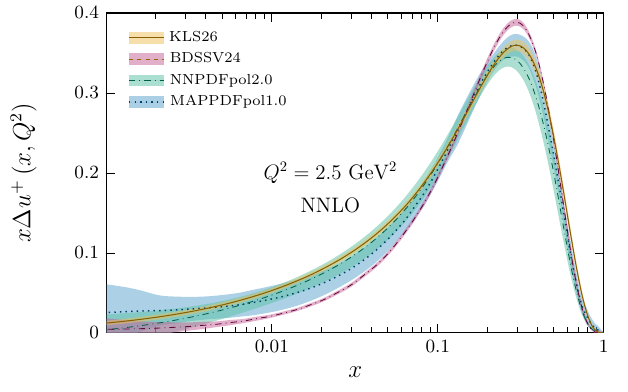}
\includegraphics[width=0.38\textwidth]{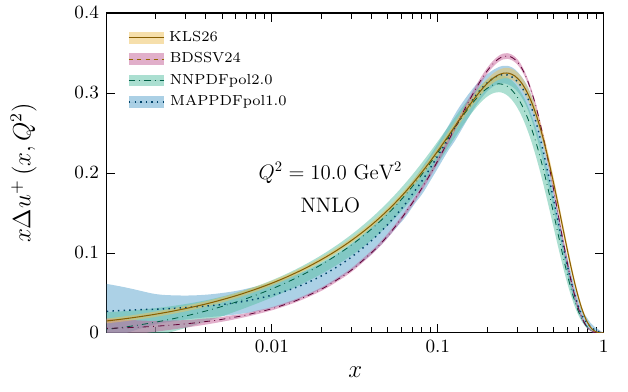}

\vspace{0.2cm}

\includegraphics[width=0.38\textwidth]{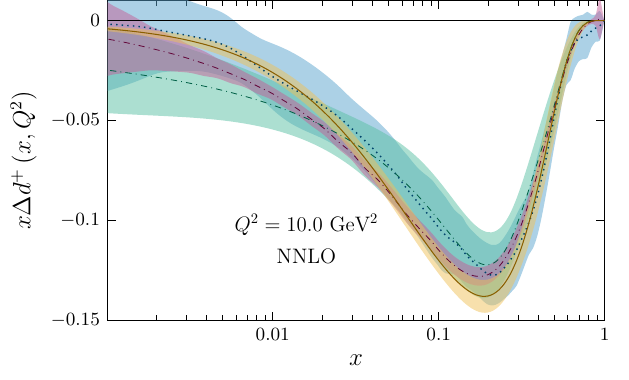}
\includegraphics[width=0.38\textwidth]{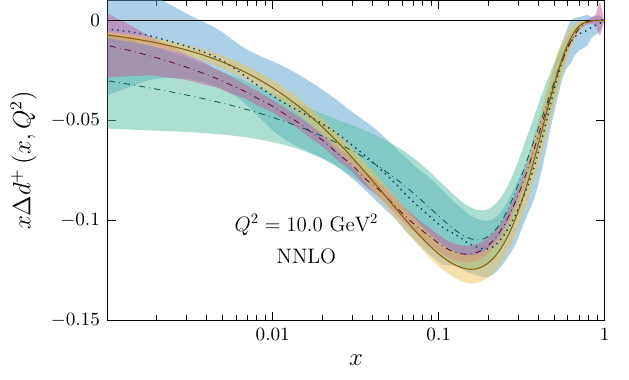}

\vspace{0.2cm}

\includegraphics[width=0.38\textwidth]{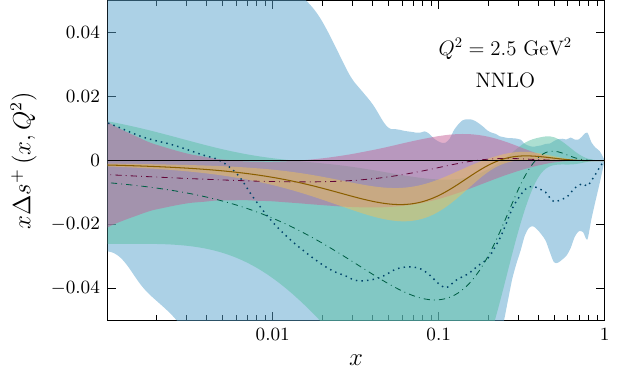}
\includegraphics[width=0.38\textwidth]{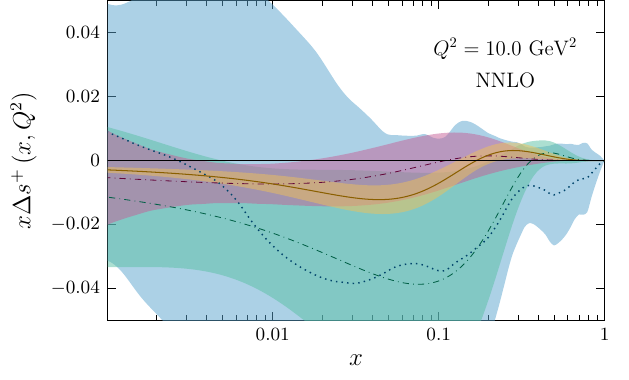}

\vspace{0.2cm}

\includegraphics[width=0.38\textwidth]{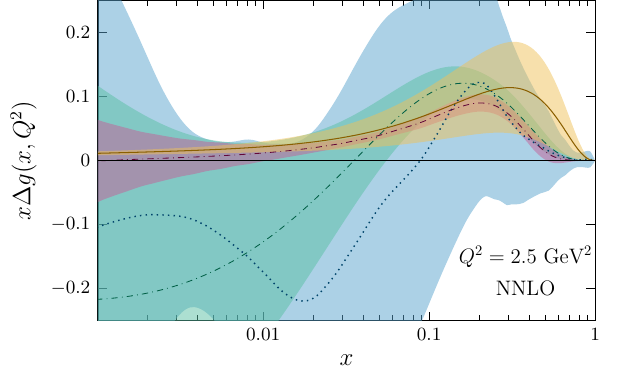}
\includegraphics[width=0.38\textwidth]{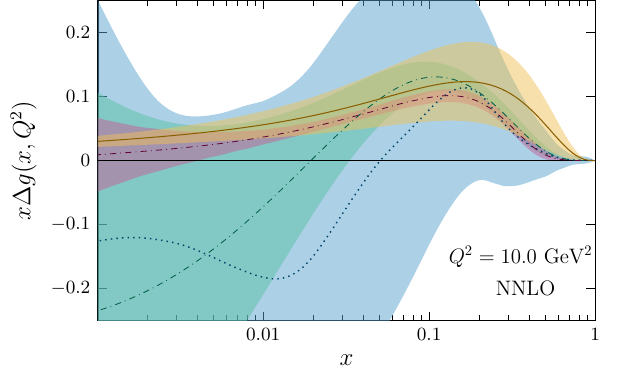}

\caption{
Polarized parton distribution functions obtained from the
present NNLO QCD analysis using additive higher-twist corrections, compared with
the BDSSV24, NNPDFpol2.0, and MAPPDFpol1.0 results.
From top to bottom, the panels show
$x\Delta u^{+}(x,Q^2)$,
$x\Delta d^{+}(x,Q^2)$,
$x\Delta s^{+}(x,Q^2)$,
and
$x\Delta g(x,Q^2)$.
The left panels correspond to
$Q^2=2.5~\mathrm{GeV}^2$,
while the right panels correspond to
$Q^2=10.0~\mathrm{GeV}^2$.
}
\label{fig:NNLOPDFs}
\end{figure*}

\subsection{Polarized Parton Distribution Functions at NNLO}

The polarized parton distributions obtained from the present NNLO global analysis are determined using exclusively inclusive polarized DIS data. The \texttt{KLS26} PDF sets are extracted with both additive and multiplicative higher-twist parametrizations.

In Fig.~\ref{fig:NLO_NNLO_PDFs}, we compare the polarized PDFs obtained in our present {\tt KLS26} NNLO analysis at $Q^2 = 2.5~\text{GeV}^2$ using the additive higher-twist parametrization with the KLSS21 NLO determination~\cite{Khorramian:2020gkr}. 
The four panels present $x\Delta u^+$, $x\Delta d^+$, $x\Delta s^+$, and the gluon helicity distribution $x\Delta g$ as functions of $x$.
For the valence-dominated distributions $x\Delta u^+$ and $x\Delta d^+$, the NNLO results show excellent overall agreement with the NLO predictions across the entire $x$ range, maintaining a prominent positive peak near $x \approx 0.3$ for $x\Delta u^+$ and a negative minimum around $x \approx 0.2$ for $x\Delta d^+$.
In contrast, the strange quark distribution $x\Delta s^+$ in the NNLO fit displays a noticeable shift toward less negative values at low-to-intermediate $x$ and turns slightly positive in the larger $x$ region. 
Furthermore, while the gluon density exhibits a similar qualitative shape in both orders, the NNLO central curve yields a slightly reduced peak height near $x \approx 0.3$ alongside a mildly enhanced positive behavior at small-$x$. As previously noted, the KLSS21 analysis utilized the MMHT14 NLO set~\cite{MMHT14} to enforce positivity constraints and adopted $\alpha_s(M_Z^2) = 0.120$ as the reference value for the strong coupling. Nevertheless, the inclusion of the higher-order perturbative QCD corrections remains the dominant driver behind the observed shifts between the two determinations.

A complementary view is provided in Fig.~\ref{fig:NNLOPDFs}, where the polarized parton distribution functions are shown at \(Q^2=2.5~\mathrm{GeV}^2\) and \(Q^2=10.0~\mathrm{GeV}^2\). The additive HT results, corresponding to the baseline determination {\tt KLS26}, are presented together with comparisons to the recent BDSSV24~\cite{Borsa:2024mss}, MAPPDFpol1.0\footnote{For the comparison with MAPPDFpol1.0, the DIS-only version of the analysis is considered in this work.}~\cite{Bertone:2024taw}, and NNPDFpol2.0~\cite{Cruz-Martinez:2025ahf} analyses.

The observed differences among the various determinations originate from the use of different types of experimental inputs and the associated kinematic selections adopted in each global analysis. In particular, while the present fit relies exclusively on inclusive polarized DIS data, other global determinations also include additional processes such as SIDIS and, in some cases, polarized proton-proton scattering, which provide complementary but differently weighted constraints on the partonic flavor separation. 

Furthermore, as shown in Fig.~\ref{fig:NNLOPDFs}, the implementation of different kinematic cuts, particularly in \(Q^2\) and \(W^2\), leads to variations in the accessible phase space and consequently in the sensitivity to different \(x\)-regions of the PDFs. The treatment of HT effects, either explicitly included or effectively absorbed depending on the analysis, introduces additional differences in the description of the low-\(Q^2\) region, where non-perturbative contributions become more relevant.
These combined effects directly impact both the central values and the uncertainty estimates of the extracted distributions, leading to the observed spread among different global determinations.

Despite these methodological differences, a general level of consistency is observed across all determinations. The \(x\Delta u^{+}\) distribution shows strong agreement over the full \(x\)-range, with all fits remaining compatible within uncertainties. For \(x\Delta d^{+}\), slightly larger deviations are observed in the intermediate-\(x\) region, although consistency is still maintained within the quoted error bands.

The \(x\Delta s^{+}\) distribution exhibits larger uncertainties due to the limited experimental constraints, leading to a broader spread among different global analyses. The polarized gluon distribution \(x\Delta g\) shows the largest sensitivity to both theoretical assumptions and indirect experimental constraints, with more pronounced differences at small \(x\).

Finally, in Fig.~\ref{fig:NNLOPDFs} the comparison between \(Q^2=2.5~\mathrm{GeV}^2\) and \(Q^2=10.0~\mathrm{GeV}^2\) demonstrates a stable QCD evolution behavior, indicating that the extracted PDFs are robust against scale variation within the NNLO framework.

\subsection{Impact of Releasing the Axial Charge Constraints}
\label{subsec:free_axial_charges}

As an additional stability test of the NNLO analysis, we
perform independent fits in which the nonsinglet axial charges
$a_3$ and $a_8$ are treated as free parameters rather than
being fixed to their experimental values. This analysis is
carried out separately for the additive and multiplicative
higher-twist parametrizations.

For the additive higher-twist parametrization, the minimization
yields
\[
a_3 = 1.253 \pm 0.043,
\qquad
a_8 = 0.780 \pm 0.274,
\]
with a total $\chi^2=504.61$. For the multiplicative
higher-twist parametrization, we obtain
\[
a_3 = 1.225 \pm 0.047,
\qquad
a_8 = 0.838 \pm 0.315,
\]
with a total $\chi^2=501.61$.

In both cases, the differences in the total $\chi^2$ between
the additive and multiplicative higher-twist
implementations remain very small, indicating that both
parametrizations provide an equally stable description of
the present polarized inclusive DIS data.

\begin{figure}[!tbp]
\centering
\includegraphics[width=0.35\textwidth]{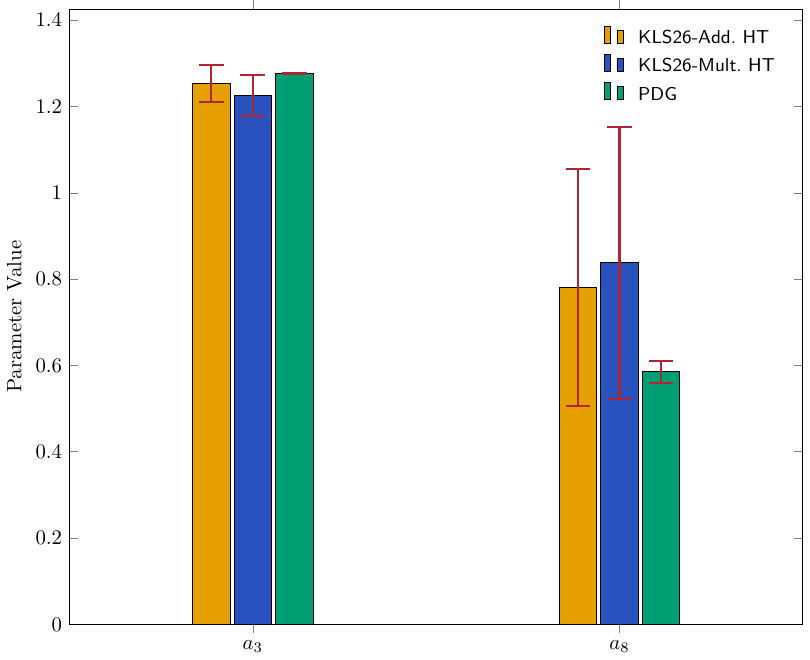}
\caption{
Comparison of the axial charges $a_3$ and $a_8$ extracted in
the {\tt KLS26} additive and multiplicative higher-twist frameworks
with the corresponding PDG averages. The error bars denote the
one-sigma uncertainties. The values obtained in
the two {\tt KLS26} fits are compatible with the PDG determinations
within the quoted uncertainties, while the fitted value of
$a_8$ exhibits a considerably larger uncertainty.
}
\label{fig:a3a8_comparison}
\end{figure}

These results may be compared with the Particle Data Group
averages~\cite{PDG},
\[
a_3 = 1.2756 \pm 0.0013,
\qquad
a_8 = 0.585 \pm 0.025,
\]
which are used as external inputs in the default fits with fixed
$a_3$ and $a_8$. The values of $a_3$ extracted from polarized
inclusive DIS are compatible with the standard experimental determination. The extracted values of $a_8$ also remain
compatible within their substantially larger uncertainties,
reflecting the weaker sensitivity of inclusive polarized DIS
data to the octet axial charge and, consequently, to the
polarized strange-quark sector. Figure~\ref{fig:a3a8_comparison} provides a direct comparison of the fitted axial charges in both the additive and multiplicative HT frameworks with the corresponding PDG values.

Most importantly, allowing $a_3$ and $a_8$ to vary freely has
only a negligible effect on the overall fit quality relative to
the default fits. In the additive higher-twist analysis, the
total $\chi^2$ changes from $\chi^2=504.41$ in the fixed axial charge
fit to $\chi^2=504.61$ in the  free axial charge fit. In the
multiplicative case, it changes from $\chi^2=501.77$ to
$\chi^2=501.61$. These small variations indicate that the
present polarized inclusive DIS data do not strongly prefer a
specific fixed axial charge input and that the external
constraints on $a_3$ and $a_8$ do not artificially control the
quality of the fit.

\begin{figure*}[!htbp]
\centering
\includegraphics[
  width=0.90\textwidth
]{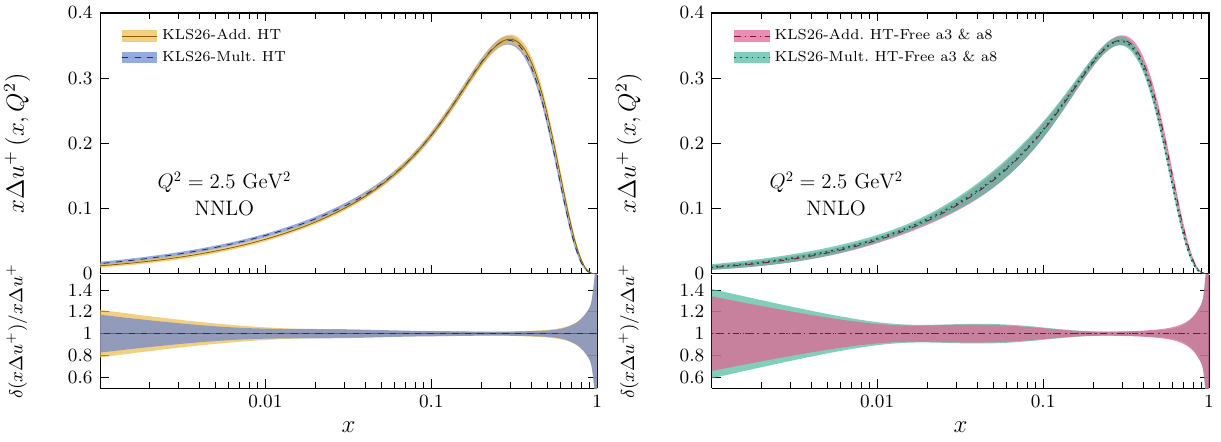}
\\
\includegraphics[
  width=0.90\textwidth
]{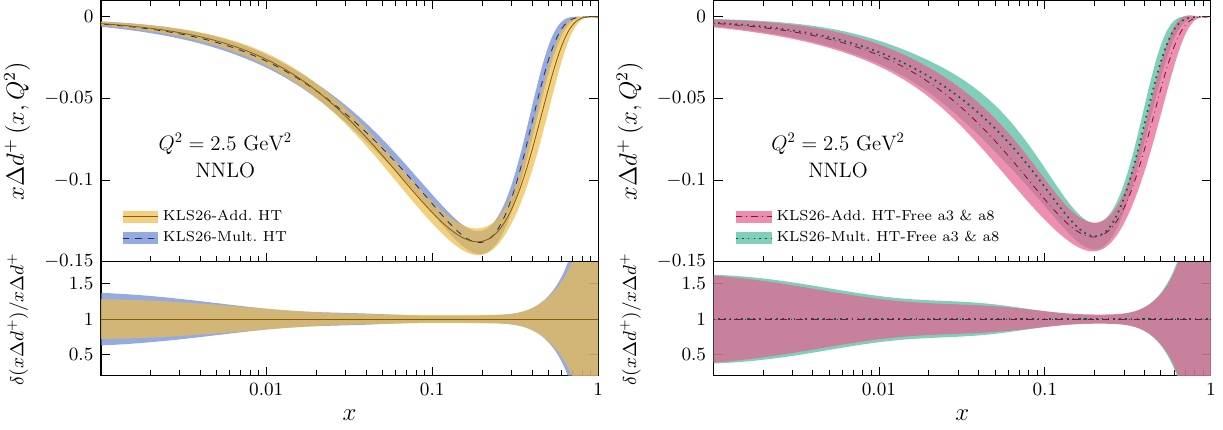}
\\
\includegraphics[
  width=0.90\textwidth
]{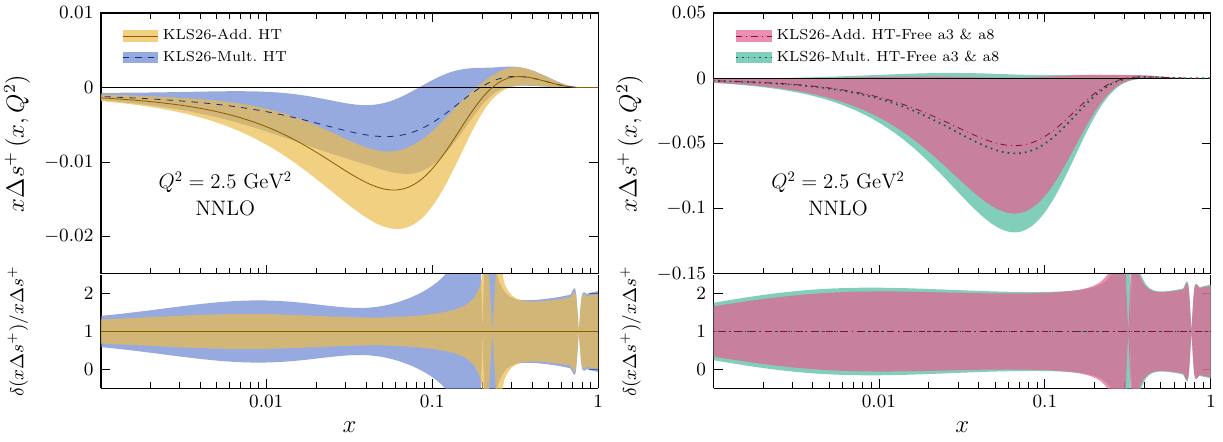}
\\
\includegraphics[
  width=0.90\textwidth
]{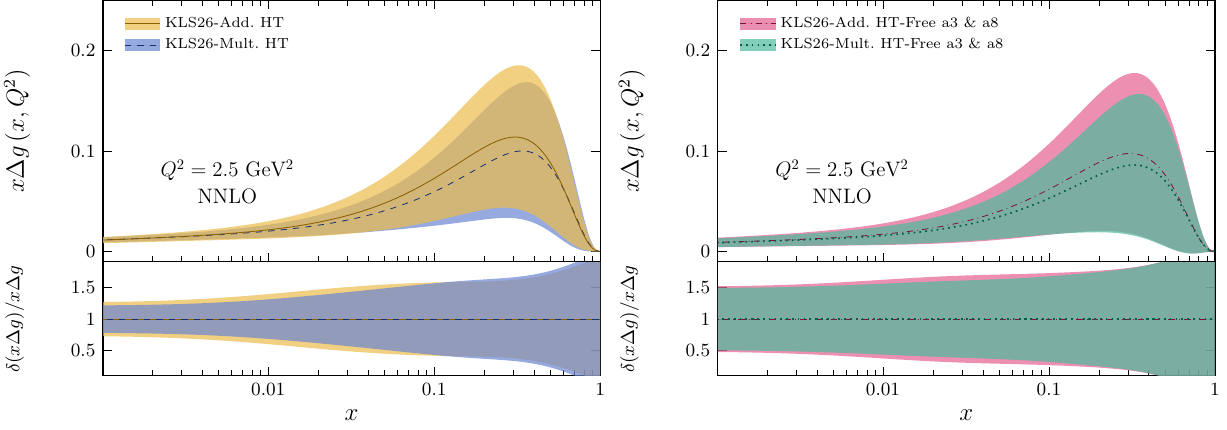}
\caption{
Polarized parton distributions at
$Q^2=2.5~\mathrm{GeV}^2$ for the additive and
multiplicative higher-twist implementations. The left panels
correspond to the baseline scenario with fixed axial charges
$a_3$ and $a_8$, while the right panels show the results
obtained when $a_3$ and $a_8$ are treated as free parameters.
In each case, the additive and multiplicative higher-twist
fits are displayed for comparison. The lower panels show the
corresponding relative PDF uncertainties,
$\delta(x\Delta q^+)/x\Delta q^+$, evaluated with respect to
the central value of each individual fit.
}
\label{fig:SU3_HT_PDFs}
\end{figure*}

Figure~\ref{fig:SU3_HT_PDFs} compares the polarized PDFs
obtained with fixed and free axial charges for both
higher-twist implementations. The upper panels show the
absolute distributions, while the lower panels display the
corresponding relative PDF uncertainties,
${\delta\!\left(x\Delta q^+\right)}/
     {x\Delta q^+}.
$
The left panels correspond to the baseline fits with fixed
$a_3$ and $a_8$, whereas the right panels show the results
obtained when both axial charges are treated as free
parameters. This organization permits a direct comparison of
the effects associated with the higher-twist prescription and
the axial charge constraints.

The $\Delta u^+$ distribution remains highly stable when the
axial charge constraints are released, with only negligible
changes in its central value relative to the baseline fit. A
similar pattern is observed for $\Delta d^+$, although its
uncertainty band increases more noticeably in the
free-$\{a_3,a_8\}$ fits, indicating a mild sensitivity to the
removal of the external constraints.

The relative-uncertainty panels further show that the
differences generated by the additive and multiplicative
higher-twist treatments remain small over most of the
kinematic range. By contrast, releasing the axial charge
constraints produces a more visible enlargement of the PDF
uncertainties, particularly in the small- and
intermediate-$x$ regions. This effect is stronger for
$\Delta d^+$ than for $\Delta u^+$. The extracted
light-quark distributions are therefore more sensitive to the
treatment of the axial charges than to the specific
higher-twist parametrization, although both distributions
remain comparatively well constrained.

The most significant effect is observed in the polarized
strange-quark distribution, $\Delta s^+$. In both
higher-twist implementations, releasing the axial charge
constraints shifts the central distribution toward more
negative values and produces a substantial enlargement of its
uncertainty band. The relative-uncertainty panels demonstrate
that the differences between the additive and multiplicative
higher-twist treatments remain comparatively small, whereas
freeing $a_3$ and $a_8$ leads to a much larger increase in the
uncertainty over a broad range of $x$, particularly in the
small- and intermediate-$x$ regions.

This behavior reflects the central role of the octet axial
charge $a_8$ in constraining the polarized strange-quark
distribution through the axial-vector relation Eq.~\eqref{eq:a8moment}. Once this external constraint is
removed, a substantially wider range of strange-quark
solutions becomes compatible with the polarized inclusive DIS
data.

The polarized gluon distribution $\Delta g$ exhibits a
moderate sensitivity to both the higher-twist implementation
and the treatment of the axial charge constraints. In the
baseline scenario with fixed $a_3$ and $a_8$, the additive and
multiplicative fits display visible differences in the central
value, with the additive solution favoring a somewhat larger
positive gluon polarization in the intermediate- and
large-$x$ regions.

When $a_3$ and $a_8$ are treated as free parameters, the
central gluon distributions remain compatible within the
enlarged uncertainty bands. The relative PDF uncertainty is
sizeable in all four fit scenarios, reflecting the limited
direct sensitivity of inclusive polarized DIS data to the
gluon distribution. The principal consequence of releasing
the axial charge constraints is therefore an enlargement of
the gluon uncertainty rather than a systematic displacement
of its central value.

Overall, Fig.~\ref{fig:SU3_HT_PDFs} shows that the effect of
the higher-twist parametrization on the extracted polarized
PDFs is generally smaller than the effect associated with the
treatment of the axial charge constraints. The $\Delta u^+$
and $\Delta d^+$ distributions remain comparatively robust,
whereas the strongest sensitivity occurs in the polarized
strange-quark sector because of its direct connection with
the octet axial charge $a_8$.

\begin{figure}[!htbp]
\centering
\includegraphics[
  width=0.35\textwidth
]{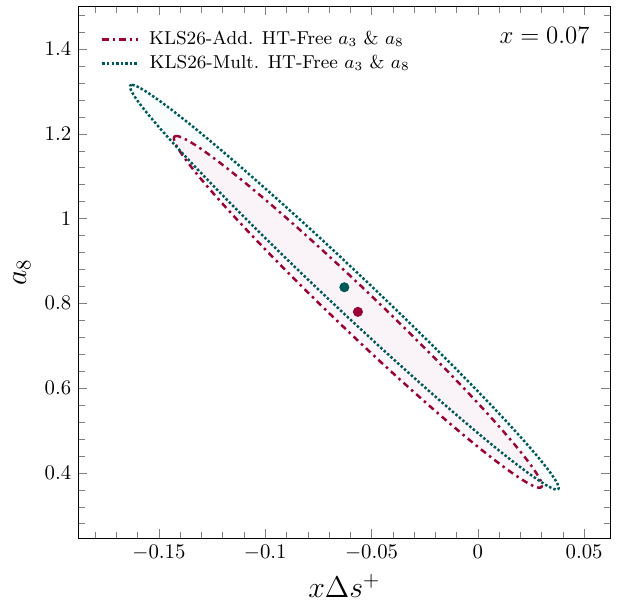}
\hfill
\includegraphics[
  width=0.35\textwidth
]{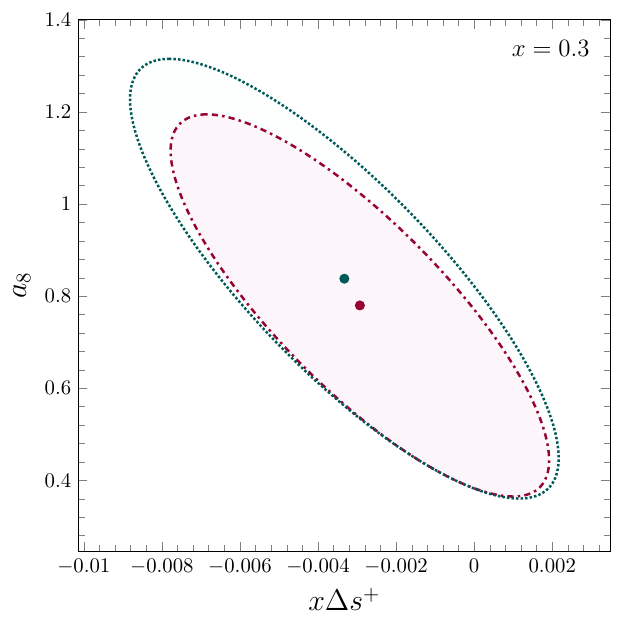}
\caption{
Representative Hessian confidence ellipses
showing the correlation between the input polarized
strange-quark distribution, $x\Delta s^+(x,Q_0^2)$, and the
fitted octet axial charge, $a_8$, for the fits in which both
$a_3$ and $a_8$ are released. The top and bottom panels
correspond to $x=0.07$ and $x=0.30$, respectively. The
ellipses and their centers denote the $68\%$ confidence
regions and the corresponding best-fit points for the
additive and multiplicative higher-twist parametrizations.
The comparison illustrates the evolution of the covariance
structure with Bjorken $x$.
}
\label{fig:xs-a8-correlation}
\end{figure}


\begin{table*}[!htbp]
\caption{
First Mellin moments of $\Delta u^+$, $\Delta d^+$,
$\Delta s^+$, and $\Delta g$ at
$Q^2=2.5~\mathrm{GeV}^2$ for the fixed- and
free-$\{a_3,a_8\}$ NNLO fits obtained with the additive and
multiplicative higher-twist parametrizations.
}
\label{tab:pdf_moments_alt}
\begin{ruledtabular}
\begin{tabular}{lcccc}
Fit scenario
&
$\Delta u^+$
&
$\Delta d^+$
&
$\Delta s^+$
&
$\Delta g$
\\
\hline
{Additive HT}
&
&
&
&
\\
\quad Fixed $a_3,a_8$
&
$0.8924 \pm 0.0352$
&
$-0.3822 \pm 0.0371$
&
$-0.0370 \pm 0.0169$
&
$0.3442 \pm 0.1878$
\\
\quad Free $a_3,a_8$
&
$0.8923 \pm 0.0525$
&
$-0.3598 \pm 0.0609$
&
$-0.1234 \pm 0.1268$
&
$0.2881 \pm 0.2139$
\\
\hline
{Multiplicative HT}
&
&
&
&
\\
\quad Fixed $a_3,a_8$
&
$0.9087 \pm 0.0352$
&
$-0.3654 \pm 0.0346$
&
$-0.0202 \pm 0.0176$
&
$0.3083 \pm 0.1637$
\\
\quad Free $a_3,a_8$
&
$0.8942 \pm 0.0587$
&
$-0.3299 \pm 0.0567$
&
$-0.1365 \pm 0.1474$
&
$0.2580 \pm 0.1830$
\\
\end{tabular}
\end{ruledtabular}
\end{table*}

To obtain a more direct understanding of this behavior, we
further examine the correlation between the local value of the
polarized strange-quark distribution,
$x\Delta s^+(x,Q_0^2)$, and the fitted octet axial charge,
$a_8$. Figure~\ref{fig:xs-a8-correlation} displays representative
$68\%$ C.L. Hessian confidence ellipses at two characteristic
Bjorken-$x$ values, $x=0.07$ and $x=0.30$, for the fits in which
both $a_3$ and $a_8$ are released. These representative $x$ values illustrate how both the orientation and the strength
 of the covariance evolve with $x$, from the intermediate-$x$ region, where polarized DIS data provide significant sensitivity
  to the strange-quark helicity distribution, to the larger-$x$ region.

Unlike the axial charge sum rules, which constrain integrated
moments of the polarized PDFs, the covariance ellipses probe
the local response of the strange-quark helicity distribution
at fixed  $x$. They therefore provide a geometric
visualization of how the information associated with the
axial charge constraints propagates through the global
Hessian covariance matrix and influences the extracted local
$x$-dependent strange-quark distribution.

A clear negative correlation between $x\Delta s^+$ and
$a_8$ is observed at the representative intermediate-$x$
value $x=0.07$, where polarized DIS data provide significant
sensitivity to the strange-quark helicity distribution.
This behavior follows directly from the octet combination of
Eq.~(\ref{eq:a8moment}). Because the light-quark moments are
already tightly constrained by polarized DIS data, an increase
in $a_8$ is accommodated primarily through a more negative
strange-quark contribution in order to satisfy the octet
axial charge relation. This explains the observed negative
correlation between $a_8$ and $x\Delta s^+$, reflected in the
negatively sloped covariance ellipses.

The comparison between the intermediate-$x$ point,
$x=0.07$, and the larger-$x$ point, $x=0.30$, demonstrates
that both the strength and the geometry of the covariance
structure evolve with  $x$. This behavior shows that an
integrated axial charge constraint does not affect all regions
of the local strange-quark distribution uniformly. Instead,
the information imposed by the $a_8$ constraint is
redistributed over $x$ through the PDF parametrization and the
full Hessian covariance matrix. Although the negative
correlation remains clearly visible in
Fig.~\ref{fig:xs-a8-correlation}, the broader and less
elongated covariance ellipse at larger $x$ indicates that the
local response of $x\Delta s^+$ to variations of $a_8$
depends on  $x$.

For completeness, we have also examined the analogous
covariance ellipses between $x\Delta s^+$ and the triplet
axial charge $a_3$. These correlations are considerably
weaker and are therefore not shown. This is expected because
$a_3$ constrains the nonsinglet light-quark combination
through Eq.~(\ref{eq:a3moment}) and does not directly involve
the strange-quark helicity distribution. Any correlation
between $x\Delta s^+$ and $a_3$ therefore arises only
indirectly through the simultaneous fit and the global
covariance matrix.

The covariance analysis therefore provides a natural
explanation for the substantial increase in the uncertainty of
$x\Delta s^+$ once the axial charge constraints are relaxed.
More importantly, it demonstrates geometrically that the
dominant covariance structure involves the local
strange-quark helicity distribution and the octet axial
charge $a_8$, whereas the corresponding correlation with
$a_3$ remains weak and only indirect. The comparison of the
two representative $x$ values further shows that the
propagation of information from the axial charge constraints
through the Hessian covariance matrix is intrinsically
$x$ dependent.

To complement the local PDF-level and covariance analyses,
Table~\ref{tab:pdf_moments_alt} summarizes the corresponding
first Mellin moments of $\Delta u^+$, $\Delta d^+$,
$\Delta s^+$, and $\Delta g$ at
$Q^2=2.5~\mathrm{GeV}^2$. Results are provided for the
baseline and free-$\{a_3,a_8\}$ fits in both the additive and
multiplicative higher-twist frameworks. The table offers a
compact quantitative measure of how releasing the
axial charge constraints propagates into the integrated
helicity densities.

\begin{figure}[!htbp]
\centering
\includegraphics[
  width=0.95\columnwidth
]{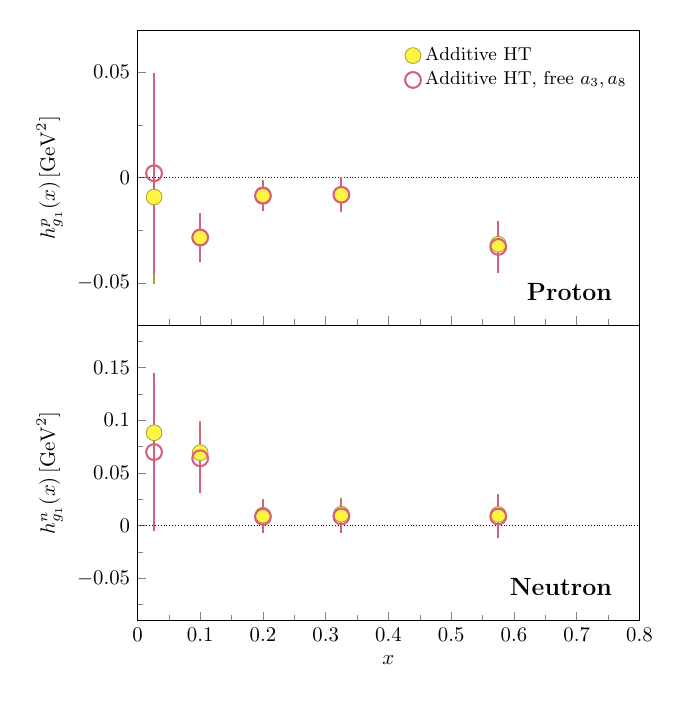}
\includegraphics[
  width=0.95\columnwidth
]{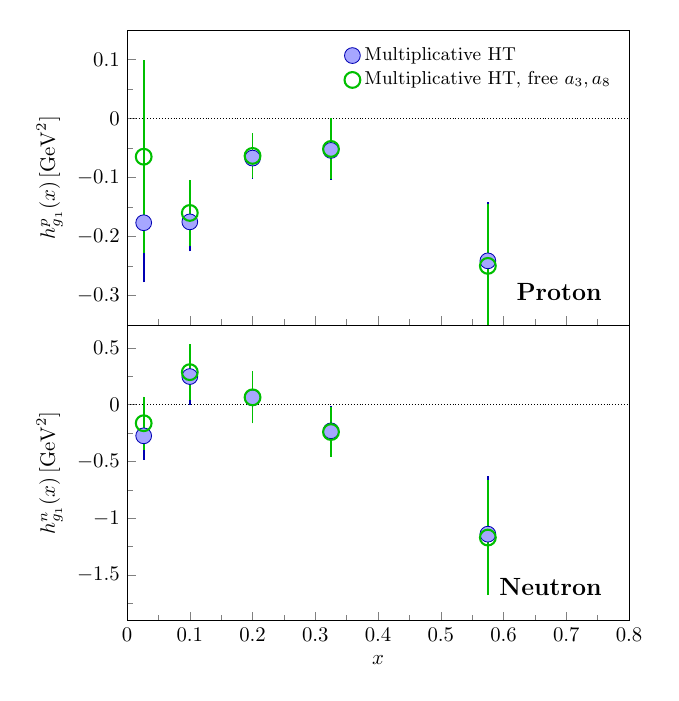}
\caption{
Extracted higher-twist contributions to the polarized proton
and neutron structure functions in the NNLO analysis with the additive (top) and
multiplicative (bottom) higher-twist parametrization. Filled markers
correspond to the baseline fit with fixed $a_3$ and $a_8$,
while open markers show the results obtained when both axial
charges are treated as free parameters.
}
\label{fig:HT_Add_Mult}
\end{figure}

The first moments of the dominant light-quark combinations
$\Delta u^+$ and $\Delta d^+$ remain comparatively stable
across the four fit scenarios. In particular, the first
moment of $\Delta u^+$ is nearly unchanged when $a_3$ and
$a_8$ are released. The changes in $\Delta d^+$ are somewhat
larger but remain compatible within the enlarged uncertainties
of the free-$\{a_3,a_8\}$ fits. These results confirm at the
level of integrated moments that the light-quark sector
remains well constrained by the polarized inclusive DIS data.

By contrast, the most pronounced effect is observed in the
polarized strange-quark moment. In the additive fit,
$\Delta s^+$ changes from
$
-0.0370 \pm 0.0169
$
in the fixed axial charge analysis to
$
-0.1234 \pm 0.1268
$
when the axial charges are released. Similarly, in the
multiplicative fit, it changes from
$
-0.0202 \pm 0.0176
$
to
$
-0.1365 \pm 0.1474.
$
Thus, in both higher-twist frameworks, releasing $a_3$ and
$a_8$ shifts the strange-quark moment toward more negative
values and produces a dramatic increase in its uncertainty.
This pattern is consistent with both the local PDF comparison
in Fig.~\ref{fig:SU3_HT_PDFs} and the covariance ellipses in
Fig.~\ref{fig:xs-a8-correlation}.

The gluon first moment exhibits a milder dependence on the
axial charge treatment. In the additive analysis, $\Delta g$
changes from $0.3442\pm0.1878$ to $0.2881\pm0.2139$, while
in the multiplicative analysis it changes from
$0.3083\pm0.1637$ to $0.2580\pm0.1830$. Although the gluon
uncertainties also increase, the effect remains substantially
smaller than in the strange-quark sector. The dominant
sensitivity to the removal of the axial charge constraints is
therefore concentrated in $\Delta s^+$.

Overall, Table~\ref{tab:pdf_moments_alt} confirms the
stability pattern inferred from the local PDF analysis. The
integrated $u^+$ and $d^+$ helicity densities remain robust,
the gluon moment displays only moderate variations, and the
strongest response occurs in the polarized strange-quark
moment because of its direct relation to the octet axial
charge.

Finally, we examine whether releasing the axial charge
constraints significantly affects the extracted higher-twist
contributions. Figure~\ref{fig:HT_Add_Mult} compare the results obtained, for protons and neutrons, with fixed and
free $a_3$ and $a_8$ for the additive and multiplicative
higher-twist implementations, respectively.

The higher-twist contributions obtained in the fixed- and
free-axial charge analyses remain mutually compatible within
their uncertainties. This stability indicates that the
principal effect of releasing $a_3$ and $a_8$ is absorbed by
the polarized PDF sector, particularly by the strange-quark
distribution, rather than by a substantial reorganization of
the fitted higher-twist contributions.

\subsection{Summary of the NNLO Results}

The NNLO analysis presented in this work provides a
consistent overall description of the world polarized
inclusive DIS data over a broad kinematic range, including
the preasymptotic region where target-mass and higher-twist
effects play an important role. Both additive and
multiplicative higher-twist parametrizations lead to very
similar global fit qualities and reproduce the available
polarized DIS measurements with comparable accuracy.

The extracted polarized parton distributions are found to
be generally stable under variations of the higher-twist
treatment. In particular, the $u^{+}$ and $d^{+}$ helicity
distributions remain highly robust across the different
fit scenarios considered in this work. 
The polarized strange-quark distribution exhibits the
strongest sensitivity to the axial charge assumptions,
whereas its dependence on the higher-twist
parametrization is less pronounced than its sensitivity
to the axial charge assumptions. The
polarized gluon distribution exhibits a moderate
sensitivity to both the higher-twist implementation and
the treatment of the axial charge constraints, primarily
through changes in its uncertainty band rather than
systematic shifts of its central value.

A further stability test was performed by releasing the two axial nonsinglet parameters, \(a_3\) and \(a_8\), and fitting them directly to the polarized inclusive DIS data. This
exercise shows that the overall fit quality changes only
marginally relative to the default fits with fixed axial
charges. 
The resulting $u^{+}$ and $d^{+}$ distributions retain
nearly unchanged central values, although the
uncertainty of these distributions increases
moderately after the axial charge constraints are
freed, whereas the largest response
is observed in the polarized strange-quark sector, as
expected from the direct role of $a_8$ in constraining the
strange-quark combination through the octet axial-vector
sum rule. The extracted higher-twist coefficients also
remain generally stable under this variation, although a
somewhat larger sensitivity is observed in the
multiplicative parametrization, particularly in the
neutron sector.

Taken together, these results indicate that the main
phenomenological conclusions of the present NNLO analysis
are not driven by a particular choice of higher-twist
ansatz. On the other hand, the polarized strange quark distribution, not surprisingly, is more sensitive to the choice of axial charge input values.

While the detailed numerical values of the higher-twist
coefficients remain model dependent, the extracted
helicity PDFs exhibit a substantially higher degree of
stability, with the strongest sensitivity observed for
the polarized strange-quark distribution.

These observations provide a consistency check of the NNLO
polarized PDF determination and support the robustness of
the present framework for future accurate studies  of the
spin structure of the nucleon.

\section*{Availability of Polarized PDF Sets}
\label{sec:LHAPDF}
The polarized parton distribution function (PDF) sets determined in this NNLO analysis are publicly available in the standard \textsc{LHAPDF6} format~\cite{Buckley:2014ana} at 
\begin{center}
  \small
  \url{https://github.com/Khorramian/KLS26pol-NNLO-LHAPDF}.
\end{center}
The released grids include the central fit together with the complete set of Hessian uncertainty eigenvectors.

For comprehensive phenomenological applications, we provide grid sets corresponding to both additive and multiplicative higher-twist parametrizations, as well as fits treating the axial charges ($a_3, a_8$) as free parameters. These sets are derived from the NNLO QCD analysis of polarized inclusive deep-inelastic scattering data presented in this study and are intended for standardized use in high-energy physics predictions.

\section*{Acknowledgements}
We are particularly grateful to V.~Bertone for the
development of \texttt{APFEL++} and for many helpful
discussions, technical suggestions, and support
throughout this work. We also thank A.~Ciefa for kindly
making available the MAPPDFpol DIS sets used in
the comparisons presented in this paper. A.K. gratefully
acknowledges the hospitality of the CERN TH-PH Division,
where part of this work was carried out.
J.S.\ is supported by a Postdoctoral Fellowship from Semnan University under Contract No.~20251144.

\end{document}